\documentclass[preprint, 
superscriptaddress,floatfix,aps,prb,showpacs]{revtex4}
\usepackage{graphicx} 

\begin{document}

\newcommand{\ket}[1]{|#1\rangle}              
\newcommand{\bra}[1]{\langle #1|}             
\newcommand{\braket}[2]{\langle #1|#2\rangle} 
\def\la{\langle}
\def\ra{\rangle}
\def\wh{\widehat}
\newcommand{\beq}{\begin{equation}}
\newcommand{\eeq}{\end{equation}}
\newcommand{\beqa}{\begin{eqnarray}}
\newcommand{\eeqa}{\end{eqnarray}}
\newcommand{\intf}{\int_{-\infty}^\infty}
\newcommand{\into}{\int_0^\infty}

\title{Quantum kinetic energy densities: An operational approach}
\author{J. G. Muga}
\affiliation{* Departamento de Qu\'\i mica-F\'\i sica, UPV-EHU, Apdo. 
644, Bilbao, Spain}
\author{D. Seidel}
\affiliation{Institut f\"ur Theoretische Physik, Universit\"at G\"ottingen,
Friedrich-Hund-Platz~1, 37077 G\"ottingen, Germany}
\author{G. C. Hegerfeldt}
\affiliation{Institut f\"ur Theoretische Physik, Universit\"at G\"ottingen,
Friedrich-Hund-Platz~1, 37077 G\"ottingen, Germany}

\begin{abstract}
We propose and investigate a procedure to measure, at least in
principle, a positive 
quantum version of the  local kinetic energy density. This procedure 
is based, under certain idealized limits, on the detection rate of photons
emitted by moving atoms which are excited by a localized laser beam.
The same type of experiment, but in different limits, can also provide
other non positive-definite versions of the kinetic energy
density. A connection with quantum arrival time distributions is discussed.  
\end{abstract}
\maketitle

\section{Introduction}
%
%
To obtain an expression for the local density of a quantum observable
 not  diagonal in coordinate representation, one may look for
guidance to the  corresponding classical case. For a classical 
dynamical variable, $A(q,p)$, in position-momentum phase space
its local density, $\alpha_A(x)$, is simply obtained by
\beqa
 \alpha_A(x)&=& \int dp \rho(x,p) A(x,p)\\ \nonumber
                      &=& \int dq dp \rho(q,p) \delta(x-q) A(q,p)
\eeqa
where $\rho(q,p)$ is the phase space density.
To quantize this expression one can use
\beq
\delta(x-\widehat{x})=|x\ra\la x| 
\eeq
and consider, for a point $x$,  the operator
$\widehat{A}(x)=\widehat{A}|x\ra\la x|$, or rather one of 
its many symmetrizations, as a quantum density for the observable
$\widehat{A}$. For a given state $|\psi \ra$, the expectation value  $\la 
\psi|\widehat{A}(x)|\psi \ra$  would then
be a candidate for the value of the local density at the
point $x$ of the observable $\widehat{A}$. If  $\widehat{A}$  is not 
diagonal in coordinate representation so that it does not commute with
$|x\ra\la x|$ , there are  infinitely many  ``combinations'' 
(orderings) to construct a quantum density,\cite{CL02,MPS98} for example, 
\beqa
\widehat{A}^{1/2}|x\ra\la x|\widehat{A}^{1/2}
\\
\frac{1}{2}\left(\widehat{A}|x\ra\la x|+|x\ra\la x|\widehat{A}\right)
\\
\frac{1}{2}\left(\widehat{A}^{1/2}|x\ra\la x|\widehat{A}^{1/2}\right)
+\frac{1}{4}\left(\widehat{A}|x\ra\la x|+|x\ra\la x|\widehat{A}\right)
\eeqa 
The noncommutativity of two observables does not mean that there is
only one ``true'' symmetrization of their product.   
Different symmetrizations  may have a perfectly 
respectful  status as physically observable and measurable quantities, and 
different orderings may be associated with latent properties 
that may be realized via different 
experimental measurement procedures. They may also be related more
indirectly to observables and yet carry valuable physical
information. An example of this non-uniqueness due to different
symmetrizations is the arrival time of a quantum particle at a
particular position. Classically the distribution 
of arrival times would be the flux for particles 
moving in one direction, but quantum
mechanically 
different quantizations have been proposed.\cite{ML00,MSE02}

Another important example of this quantum non-uniqueness for
a single classical quantity is the kinetic energy density,
\cite{Cohen79,Cohen84,Robinett95}  in which case $\widehat{A}$ becomes the
kinetic energy operator, $\widehat{A}=\widehat{p}^2/2m$.
There is no unique definition of a quantum ``kinetic energy 
density'', in spite of the relevance of the concept in several 
fields. The Thomas-Fermi theory provides 
an early example of a possible realization and application. In density 
functional theory, it enters as one of the terms of the
energy functional to determine the electronic structure of  
atoms, molecules, solids or fermionic gases, see e.g.
Refs. \onlinecite{APN02,BZ01}.  
In this context it has been used in particular to define a local
temperature and  identify molecular sites most reactive to 
electrophilic reagents.\cite{APN02}
The kinetic energy density also plays a key role in 
partitioning molecular systems into fragments with well defined energies
and virial relations,\cite{Cohen79,BB72,Bader90} or
to define ``intrinsic shapes'' of reactants, transition states and 
products along the course of a chemical reaction.\cite{Tachibana01} 
It is moreover a basic quantity in quantum 
hydrodynamic equations.\cite{TS70,MPS98}  

In all these applications, different quantum versions of the 
kinetic energy density have been used. The most commonly found cases 
are the three quantizations considered above, or suitable
generalizations thereof. They satisfy different properties and there
have been intensive discussions
which one is best, but clearly 
they all are useful in different ways and for different purposes.
However, not so much emphasis has been placed on possible procedures
of how to measure them.


Many of these arguments and controversies already can be seen in the
simple  case of the kinetic energy density 
of a free particle in one dimension. Real, Muga and Brouard
\cite{RMB97}  studied and compared 
three versions of  an operator, $\widehat{\tau}(x)$, for the kinetic
energy density at a point $x$,
associated with different quantizations, namely 
\beqa
\widehat{\tau}^{(1)}(x)&=&\frac{\wh{p}\delta(x-\wh{x})\wh{p}}{2m},
\\
\widehat{\tau}^{(2)}(x)&=&\frac{1}{2}
\left[\frac{\wh{p}^2}{2m}\delta(x-\wh{x})+\delta(x-\wh{x})
\frac{\wh{p}^2}{2m}\right],
\\  
\widehat{\tau}^{(3)}(x)&=&\frac{1}{2}\left[\widehat{\tau}^{(1)}(x)+
\widehat{\tau}^{(2)}(x)\right]~.
\eeqa
The second operator follows from the quantization rule of Rivier.
\cite{Rivier-PR-1957} The corresponding  density $\la
\widehat{\tau}^{(2)}\ra_t$ is given by its, generally time dependent,  
expectation value and may in principle be obtained operationally by a weak
measurement of the kinetic energy
post-selected at position $x$.\cite{AV90,APRV93,Johansen2004}  
The third one, which is the average of $\la \widehat{\tau}^{(1)}\ra_t$ and 
$\la \widehat{\tau}^{(2)}\ra_t$,  corresponds to Weyl's quantization rule. 
An indirect  way to measure 
$\la \widehat{\tau}^{(3)}\ra_t$ for free motion was 
described by Johansen,\cite{Johansen98} who noticed that the second
time derivative of 
the expectation value of $|\widehat{x}-x|$ is proportional
to $\la \widehat{\tau}^{(3)}\ra_t$. 

In this paper we will provide an operational interpretation of the
first expression which, incidentally, is the only positive one
among the three, and in fact among a much broader family of
quantizations.\cite{APN02}
We will use for this end a simple model, originally
devised to study  time of arrival measurements.\cite{DEHM02}
The basic physical idea is to send atoms in their ground state through
a region illuminated by a perpendicular laser beam and to measure
the resulting
fluorescence rate of photons.   

\section{Kinetic energy density, fluorescence and atomic absorption
  rate in an imaginary   potential barrier} 

The description of photon fluorescence from moving atoms, which are excited
by a localized laser beam, is based on the quantum jump approach
\cite{Hegerfeldt93, DCM92, Carmichael93} and has been discussed in detail 
elsewhere.\cite{DEHM02, DEHM03, HSM03} We will only 
summarize the results which are relevant for the present investigation  
and  assume a ``lambda'' configuration of three atomic levels 
in which the laser couples levels 1 and 2 with Rabi frequency $\Omega$,
whereas level 2 decays with inverse life time $\gamma$ 
predominantly and irreversibly to a ground (sink) state 3.\cite{OABSZ96} 
For a laser on resonance with the atomic transition, atoms for which
no photon is detected (``undetected atoms'')
 are  governed by the following effective Hamiltonian 
\begin{equation}
\widehat{H} = {\widehat{p}}^2/2m + \frac{\hbar}{2} \Omega(\widehat{x})
\left\{ |2\rangle\langle 1| +
|1\rangle\langle 2|\right\}
 - \frac{i}{2} \hbar \gamma |2\rangle\langle 2|~,
\label{Hami2}
\end{equation}
where $\Omega(x)$ is the position dependent Rabi frequency. 
The evolution of the wave function for undetected atoms simplifies to 
a one channel Schr\"odinger equation if $\hbar\gamma$ is large  
compared to the kinetic energy,  with a complex imaginary potential
\cite{RDNMH04} 
\beq
U(\widehat{x})=-i\hbar\frac{\Omega(\widehat{x})^2}{2\gamma}.
\eeq
In this ``low saturation regime'' the undetected atoms, whose number
is proportional to the norm-squared of the wave-function, are 
in the ground state  most of the time, and  the temporal probability
density for a photon detection (``detection rate'') is given by the
decrease of the number of undetected atoms,  
\beq
\Pi(t)=-d\la \psi(t)|\psi(t) \ra/dt=-\frac{2}{\hbar}\la \psi(t)|{\rm
  Im}(U)|\psi(t) \ra~. 
\label{pit}
\eeq
This is the basic operational quantity obtained in the modeled experiment. 
We will relate it, or its normalized version
\beq\label{pin}
\Pi_N(t)=\frac{\Pi(t)}{\int dt\,\Pi(t)}
\eeq
to ideal quantities for the
freely moving  atom, unperturbed by the laser, by taking certain limits.

First, we consider a square laser beam profile. Then the imaginary
potential becomes  a barrier of height
\beq
V=\hbar \Omega^2/2\gamma~
\eeq
located between $x=0$
and $x=L$, and the effective Hamiltonian for undetected atoms reads
\begin{equation} \label{eff}
\widehat{H} = \frac{\widehat{p}^2}{2m} - iV\chi_{[0,L]}(\widehat{x}),
\end{equation}
where $\chi_{[0,L]}$ is 1 inside the laser illuminated region and zero
outside. The absorption, or detection, rate  is found
from Eq. (\ref{pit}) and  is given by
\begin{equation}\label{eq:abs_rate_barrier}
\Pi(t) = \frac{2V}{\hbar}\int_0^L dx\,|\psi(x,t)|^2. 
\end{equation}
To obtain the time development of a wave packet coming in from the
left, we solve first the stationary equation
\begin{equation}
\widehat{H}\phi_k = E_k \phi_k.
\end{equation}
Using  standard matching conditions, the energy eigenfunctions
$\phi_k$ in the barrier region $0 \leq x \leq L$ for a plane wave
coming in from the left with momentum  $\hbar k$  are given 
by
\begin{equation}
  \phi_k(x) = \frac{1}{\sqrt{2\pi}}\left(A_+(k) e^{iqx} + A_-(k)
    e^{-iqx} \right)
\end{equation}
with $q^2 = k^2 +2imV/\hbar^2$ and
\begin{equation}
  A_{\pm}(k) = \frac{k(q \pm k)e^{\mp iqL}}{2kq\cos(qL) -i(k^2+q^2)\sin(qL)}.
\end{equation}
Writing  the initial state as a superposition of eigenfunctions with
positive momenta, we obtain 
\begin{equation}
  \psi(x,t) = \int_0^\infty dk\,\widetilde{\psi}(k)\,e^{-i\hbar k^2 t/2m}
  \phi_k(x).
\end{equation}
Inserting this into Eq. (\ref{eq:abs_rate_barrier}) yields the
absorption rate.
It is also useful to define the auxiliary freely moving wave packet 
\begin{equation}
  \psi_f(x,t) = \frac{1}{\sqrt{2 \pi}}
\int_0^\infty dk\,\widetilde{\psi}(k)\,e^{-i\hbar k^2 t/2m}
  e^{ikx}.
\end{equation}
We will now relate the absorption rate to an ideal distribution
by going to the limit  $V\to\infty$. When $V$ is increased,
more and more atoms are reflected without being detected, 
but normalizing the result a finite distribution is
obtained in the limit,  even though the absorption
probability vanishes eventually due to total reflection.
For large $V$, $V \gg \hbar^2 k^2/2m$, one has in leading order
\begin{eqnarray}
  q &\simeq& \sqrt{2imV/\hbar^2},\nonumber\\
  A_+ &\simeq& \frac{2k}{q},\nonumber\\
  A_- &\simeq& \frac{2k}{q}e^{2iqL},\nonumber\\
  \phi_k(x) &\simeq& \frac{1}{\sqrt{2\pi}}\frac{2k}{q}\left(e^{iqx} +
    e^{iq(2L-x)}\right),\qquad 0\leq x \leq L.
\end{eqnarray}
Integrating over $x$ and neglecting the terms which  
vanish exponentially, the absorption rate becomes
\begin{eqnarray}
  \Pi(t) \simeq \frac{\hbar^2}{\pi m\sqrt{mV}} \int_0^\infty dk
\int_0^\infty dk'
  \widetilde{\psi}^*(k)\widetilde{\psi}(k')\,e^{i\hbar
    (k^2-k'^2)t/2m} kk'.
\end{eqnarray}
This expression is independent of the
barrier length $L$  as a result of the large $V$ limit,
so the same result is obtained with an  
imaginary step potential $-iV \Theta(\widehat{x})$ or with a very narrow 
barrier. 

The normalization constant is given by $\int \Pi(t) dt \simeq 2\hbar
k_0(mV)^{-1/2}$, where $k_0 = \int |\widetilde{\psi}(k)|^2 k\,dk$,
and the normalized absorption rate is
\begin{eqnarray} \label{eq:abs_rate_N}
  \Pi_N(t) \simeq \frac{\hbar}{2\pi m k_0} \int_0^\infty dk
\int_0^\infty dk'
  \widetilde{\psi}^*(k)\widetilde{\psi}(k')\,e^{i\hbar
    (k^2-k'^2)t/2m} kk'~.
\end{eqnarray}
With the freely moving wave packet $\psi_f(x,t)$  this can finally be
rewritten in the form 
\begin{equation} \label{abs}
  \Pi_N(t) \simeq
  \frac{\hbar}{mk_0}\bra{\psi_f(t)}
\widehat{k}\delta(\widehat{x})\widehat{k} \ket{\psi_f(t)}~.
\end{equation}
Now, the right hand side is just the expectation value at time t of the
kinetic energy density $\widehat{\tau}^{(1)}$ evaluated at the origin!
Thus, with $p_0=k_0\hbar$  the initial average momentum we have
obtained, in the limit $V\to \infty$,
\beq \label{tau}
\lim_{V\to\infty}\Pi_N(t)=\frac{2}{p_0}\la \widehat {\tau}^{(1)}(x=0)\ra_t~.
\eeq
Note that the averages are computed with the {\em{freely moving}} wave
function and that one obtains the kinetic energy density at an
arbitrary point $a$ by shifting the laser region, i.e. by replacing
$[0,L]$ by $[a,L+a]$ in Eq. (\ref{eff}). 

{\it Remark:} Instead of normalizing the absorption rate
of Eq. (\ref{eq:abs_rate_barrier}) by dividing by a constant one can
normalize it on the level 
of operators,\cite{BF02} which preserves it as a bilinear form. However,
in this case the result depends on the constant to which one normalizes. It has
been shown in Ref. \onlinecite{HSM03} that if the constant is chosen
as 1 then operator
normalization of Eq. (\ref{eq:abs_rate_barrier}) leads for
$V\to\infty$ to the arrival time distribution of  Kijowski.
\cite{Kijowski74} Now, since the  time integral of $\la
\widehat{\tau}^{(1)}(x=0) \ra_t$ equals $p_0/2=\hbar k_0/2$, 
it is suggestive to choose $p_0/2$ as normalization
constant. Following the approach of Ref.~\onlinecite{HSMN04} operator 
normalization of $\Pi(t)$ then leads to $\la\widehat{\tau}^{(1)}(x=0)
\ra_t$ in the limit   $V\to\infty$.

\section{Kinetic energy density from first photon measurement and
   deconvolution}
In the preceding section we had $V=\Omega^2/2\gamma$, with $\hbar\gamma$
much larger than the kinetic energy. Therefore, the limit $V\to\infty$
implies a simultaneous change of $\Omega$ and $\gamma$.
Experimentally, the Rabi frequency $\Omega$ is easy to adjust, but not
the decay rate $\gamma$. To overcome this problem we therefore describe 
a procedure that allows to keep the value of $\gamma$ fixed.

We again consider the moving-atom laser model 
but now for the limit $\Omega\to\infty$ and $\gamma = const$.
In that case, the simplified description of the evolution of the wave
function by means of the imaginary potential $U(x)$ is not feasible, and one
has to solve the full two-channel problem for the three-level atom
with the Hamiltonian in Eq. (\ref{Hami2}).
This has been done in Refs. \onlinecite{DEHM02,HSM03}, and normalizing
with a constant the resulting photon detection rate
$\Pi_N(t)$ becomes 
\begin{equation} \label{eq:first_photon_gamma}
  \Pi_N(t) \simeq \frac{\hbar}{2\pi m k_0} \int_0^\infty dk
dk' \widetilde{\psi}^*(k)\widetilde{\psi}(k')\,e^{i\hbar 
    (k^2-k'^2)t/2m}  \frac{\gamma kk'}{\gamma+i\hbar(k^2-k'^2)/m}~.
\end{equation}
For $\hbar\gamma$ large compared to the kinetic energy of the incident 
atom,
Eq. (\ref{eq:abs_rate_N}) is recovered, but for finite $\gamma$ there 
is a delay in the detection rate. This can be eliminated by means of a
deconvolution with the first-photon distribution $W(t)$ for an atom at rest.
\cite{DEHM02,HSM03} The ansatz
\begin{equation} \label{eq:deconv}
  \Pi(t) = \Pi_{id}(t)\ast W(t)
\end{equation}
yields in terms of Fourier transforms
\begin{equation} \label{eq:deconv_fourier}
  \widetilde{\Pi}_{id}(\nu) =
  \frac{\widetilde{\Pi}(\nu)}{\widetilde{W}(\nu)} 
\end{equation}
with \cite{DEHM02,KKW87}
\begin{eqnarray}
  \frac{1}{\widetilde{W}(\nu)} &=& 1 + \left(\frac{\gamma}{\Omega^2} +
    \frac{2}{\gamma} \right)i\nu + \frac{3}{\Omega^2}(i\nu)^2 +
  \frac{2}{\gamma\Omega^2}(i\nu)^3\nonumber\\
  &\simeq& 1+ \frac{2i\nu}{\gamma},\qquad \Omega\to\infty.
\end{eqnarray}
Inserting this and the Fourier transform of
Eq. (\ref{eq:first_photon_gamma}) into Eq.~(\ref{eq:deconv_fourier}),
the resulting
ideal distribution, after performing the 
inverse Fourier transform, reads
\begin{equation}
  \Pi_{id}(t) \simeq \frac{\hbar}{2\pi m k_0} \int_0^\infty dk
\int_0^\infty dk'
  \widetilde{\psi}^*(k)\widetilde{\psi}(k')\,e^{i\hbar
    (k^2-k'^2)t/2m} kk',
\end{equation}
which is the same expression as the absorption rate of
Eq.~(\ref{eq:abs_rate_N}), obtained here operationally for fixed
$\gamma$. Naturally,
\begin{equation}
  \Pi_{id}(t) \simeq \frac{2}{p_0} \la \widehat{\tau}^{(1)}(x=0) \ra_t
\end{equation}
holds as before.
\section{Connection with quantum arrival time distributions}
\label{sec:expansion} 
Here we briefly discuss a formal connection between the kinetic energy
density $\widehat{\tau}^{(1)}(x)$ given in Eq. (\ref{abs}) and 
Eq. (\ref{tau}) and the
arrival-time distribution of Kijowski\cite{Kijowski74}, 
\begin{equation} \label{eq:Kijowski}
  \Pi_K(t) = \frac{\hbar}{m} \bra{\psi_f(t)} \widehat{k}^{1/2}
  \delta(\widehat{x}) \widehat{k}^{1/2} \ket{\psi_f(t)}~,
\end{equation}
at $x=0$. For wave packets peaked around some $k_0$ in momentum space, the
operator $\widehat{k}^{1/2}$ acting on $\psi_f$ in Eq. (\ref{eq:Kijowski})
can be expanded in terms of $(\widehat{k} - k_0)$,
\begin{equation} \label{eq:Kij_expansion}
  \widehat{k}^{1/2} = k_0^{1/2} + \frac{1}{2}k_0^{-1/2}(\widehat{k}-k_0) -
  \frac{1}{8} k_0^{-3/2}(\widehat{k}-k_0)^2 +
  \mathcal{O}\left((\widehat{k}-k_0)^3\right).
\end{equation}
In the following we take $k_0$ to be the first moment of the
momentum distribution, $k_0=\int |\widetilde{\psi}(k)|^2k\,dk$.
Inserting the expansion in Eq. (\ref{eq:Kij_expansion}) into
Eq. (\ref{eq:Kijowski}) yields in zeroth order a very simple result, 
\begin{equation}
  \Pi_K(t) = v_0 |\psi_f(0,t)|^2 + \mathcal{O}(\widehat{k}-k_0)~, 
\end{equation}
i.e. the particle density times the average velocity $v_0=k_0\hbar/m$. 
To first order in $(\widehat{k} - k_0)$  one obtains the flux at $x=0$,
\begin{equation}
\Pi_K(t) = J(0,t) + \mathcal{O}\bigl((\widehat{k}-k_0)^2\bigr)~,
\end{equation}
where
\begin{equation}
  J(0,t) = \frac{\hbar}{2m} \bra{\psi_f(t)} (\widehat{k}\delta(\widehat{x}) +
  \delta(\widehat{x})\widehat{k}) \ket{\psi_f(t)}~,
\end{equation}
and to second order the expression
\begin{equation}\label{eq:Kij_expansion_2nd}
\Pi_K(t) = J(0,t) + \frac{1}{2p_0}\Delta(0,t)
+ \mathcal{O}\bigl((\widehat{k}-k_0)^3\bigr)
\end{equation}
where
\begin{equation}
\Delta(0,t)=\la \widehat{\tau}^{(1)}
(x=0)\ra_t-
\la \widehat{\tau}^{(2)}(x=0)\ra_t~.
\end{equation}
For  states with positive momentum, which we are considering here,
the first order, namely the flux, is correctly normalized to one,
and so is the second order
since the time integral over $\Delta$  is easily shown to
vanish. This difference only provides a local-in-time correction to
$J$ that   averages out globally. Its quantum nature can be further
appreciated  by the more explicit expression
\beq
\frac{1}{2p_0}\Delta =\frac{\hbar^2}{8mp_0}
\frac{\partial^2|\psi_f(0,t)|^2}{\partial x^2},
\eeq
which shows the inverse dependence on mass and momentum and the 
explicit quadratic dependence on $\hbar$.

\section{Discussion}
In Fig. 1, operational and ideal kinetic energy densities are compared 
for  a coherent superposition of two Gaussian wave
packets with different mean momenta. They are prepared in such a way
that their centers of mass arrive simultaneously at the origin. This
enhances the interference among different  momentum components and the
differences between the distributions.  
As seen in the figure, the differences between various versions  of the 
quantum kinetic energy density may be quite significant. While
$\la\widehat{\tau}^{(1)}(x)\ra_t$ is always  positive,
$\la\widehat{\tau}^{(2)}(x)\ra_t$ can become negative in classically
forbidden regions for stationary eigenstates of the Hamiltonian, a
fact that has been used  
by  Tachibana \cite{Tachibana01} to define molecular and reaction shapes. 
It is perhaps less obvious that this quantity can also be negative 
as a result of free motion dynamics, as seen in the figure.  

The main contribution of this paper has been to point out  that, under
certain idealizations and limiting conditions, 
fluorescence experiments can  lead to an operational approach to
kinetic energy densities. A positive kinetic energy density can be
obtained from the ``measured'' signal  
in a strong laser limit. With $\Delta$ the difference of the positive
density $\la \widehat{\tau}^{(1)}\ra_t$ and the density $\la
\widehat{\tau}^{2}\ra_t$ of Rivier, an interesting relation was
found for $\Delta$ with the  ideal 
time-of-arrival distribution of Kijowski and with the flux. The latter two  
can also be obtained operationally under  appropriate limits.
In  a recent review by  
Ayers, Parr and Nagy \cite{APN02} on the kinetic energy density,
one of the suggestions for future research  was the need to study and
understand this  quantity  $\Delta$ better. 
In a completely different context from the present work, and suitably
generalized to three dimensions, $\Delta$ plays a major  
role in Bader's theory to separate a molecular system into 
meaningful fragments.\cite{Cohen79,BB72,Bader90}
Note that, if $\Delta=0$, $\la \widehat{\tau}^{(3)}\ra_t$ 
becomes also equal to the other two densities considered. Therefore
this condition  implies a certain ``classicality'' or coalescence of
the multiple quantum possibilities. 
If $\Delta$ or its integral  over some volume are zero,
a fragment can be defined 
with a well defined kinetic energy and virial relations.     
It is quite striking that, in the second order expansion of the
arrival time distribution of Kijowski, $\Delta=0$ implies that 
$\Pi_K$ becomes the flux, which is, as we know, the quantity 
that plays the role of a arrival time distribution in classical mechanics.

In summary, we think that these results clarify the status, as
physically meaningful physical quantities, of several versions of the
local kinetic energy  densities, and may  stimulate experimental
research on quantum kinetic energy densities and as well as on arrival
times.

\begin{acknowledgments}
This work has been supported
by Ministerio de Ciencia y Tecnolog\'\i a, FEDER (BFM2000-0816-C03-03,
BFM2003-01003), 
UPV-EHU (00039.310-13507/2001), and Acci\'on integrada.   
\end{acknowledgments}


\newpage

FIGURE CAPTION (Fig. 1)

Comparison of kinetic energy densities at $x=0$: $\langle
    \widehat{\tau}^{(1)} 
    \rangle$ (solid), $\langle \widehat{\tau}^{(2)} \rangle$ (dashed), $\langle
    \widehat{\tau}^{(3)} \rangle$ (dotted) and the operational quantity $p_0
    \Pi_N(t)/2$, see Eqs. (\ref{pin}) and (\ref{tau}), 
    for $V=1.9\,\hbar\,\mu$s$^{-1}$, $L=0.21\,\mu$m
    (triangles) and $V=950.0\,      
    \hbar\,\mu$s$^{-1}$, $L=0.42   
    \,\mu$m (circles). 
    The initial  wave packet is a coherent combination
      $\psi=2^{-1/2}(\psi_1+\psi_2)$ of two Gaussian states for the
      center-of-mass motion of a single caesium atom that become
      separately minimal uncertainty packets (with $\Delta x_1 =
      \Delta x_2 = 0.031\,\mu$m, and average velocities $\langle
      v\rangle _1 = 18.96$ {cm/s}, $\langle
      v\rangle _2 = 5.34$ {cm/s} at $x=0$ and
      $t=2\,\mu$s). The mass is $2.2\times 10^{25}$ kg  and 
$p_0=2.67\times 10^{-26}$kg m/s. 

\newpage

$^{}$\vspace*{1cm}\\
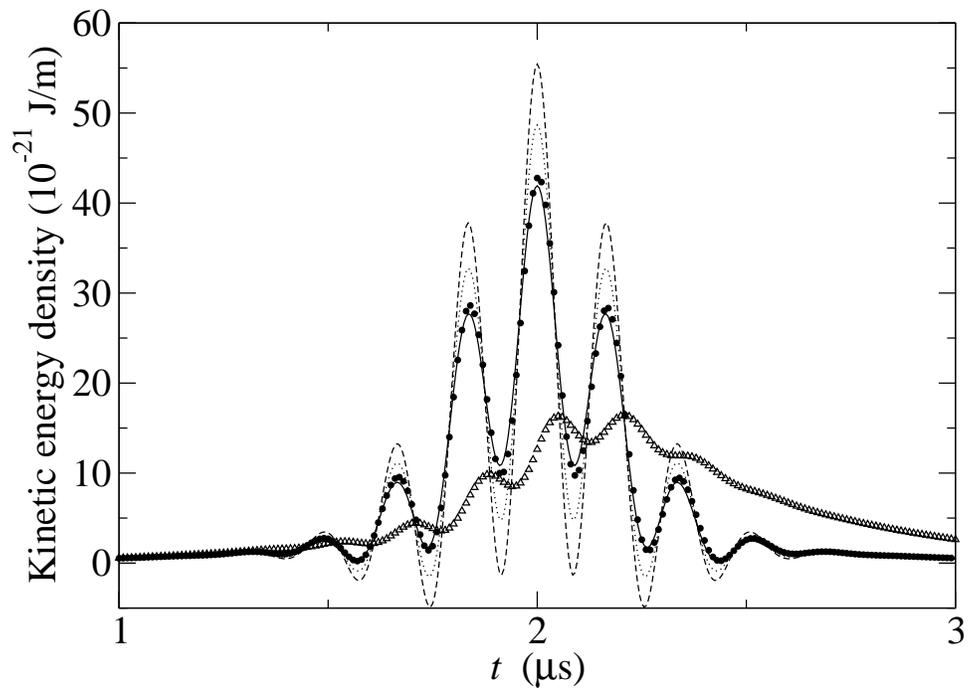
Muga, Seidel and Hegerfeldt, Figure 1

\begin{figure}
{\includegraphics[width=5in]{015515JCP1.eps}}
  \label{fig3}
\end{figure}

\end{document}